\documentclass[preprint,aps]{revtex4}
\usepackage{graphicx}
\usepackage{amssymb}
\usepackage{epstopdf}
\usepackage{amsmath}
\usepackage{listings,textcomp}
\usepackage{cancel}
\usepackage[usenames, dvipsnames]{color}
\usepackage[ pdftex, plainpages = false, pdfpagelabels, 
                 pdfpagelayout = useoutlines,
                 bookmarks,
                 bookmarksopen = true,
                 bookmarksnumbered = true,
                 breaklinks = true,
                 linktocpage=all,
                 pagebackref=false,
                 colorlinks = true,
                 linkcolor = BrickRed,
                 urlcolor  = blue,
                 citecolor = BrickRed,
                 anchorcolor = green,
                 hyperindex = true,
                 hyperfigures
                 ]{hyperref}

\newcounter{figref}

\begin{document}

\title{\href{http://necsi.edu/research/economics/options/AAPLclose.html}{The \$500.00 AAPL close: Manipulation or hedging? \\
A quantitative analysis.}}
\author{Yavni Bar-\!Yam}
\author{Marcus A.M. de Aguiar}
\author{\href{http://necsi.edu/faculty/bar-yam.html}{Yaneer Bar-\!Yam}}
\affiliation{\href{http://www.necsi.edu}{New England Complex Systems Institute} \\ 
238 Main St. Suite 319 Cambridge MA 02142, USA \vspace{2ex}}
\date{July 7, 2013; public September 3, 2013}

\begin{abstract}
Why do a market's prices move up or down? Claims about causes are made without actual information, and accepted or dismissed based upon poor or non-existent evidence. Here we investigate the price movements that ended with Apple stock closing at \$500.00 on January 18, 2013. There is a ready explanation for this price movement: market manipulation by those who sold stock options, who stood to directly benefit from this closing price. Indeed, one web commentator predicted this otherwise unlikely event publicly. This explanation was subsequently dismissed by press articles that claim that stock prices end near such round numbers based upon legitimate hedging activity. But how can we know? We show that the accepted model that points to hedging as the driving cause of prices is not quantitatively consistent with the price movement on that day. The price moved upward too quickly over a period in which the hedgersÕ position would require selling rather than buying. Under these conditions hedgers would have driven the price away from the strike price rather than toward it. We also show that a long published theory of the role of hedging is incomplete mathematically, and that the correct theory results in much weaker price movements. This evidence substantially weakens the case of those who claim hedging as cause of anomalous market price movements. The explanation that market manipulation is responsible for the final close cannot be dismissed based upon unsubstantiated, even invalid, hedging claims. Such proffered explanations shield potential illegal activity from further inquiry even though the claims behind those explanations have not been demonstrated. 
\end{abstract}
\maketitle

On Friday, January 18, 2013, Apple Inc.\ stock (AAPL) closed at exactly \$500.00 per share. This date was of particular significance in the market for side contracts for buying and selling Apple stock due to the expiration of options, with options expiring for \$1.5 billion dollars of stock based upon a price difference of just one penny in the price. 
At \$500.01 options to buy stock with a value of \$1.5 billion would not have expired, and at \$499.99 options to sell \$1.5 billion of stock would not have expired, based upon the outstanding contracts at the open of trading on that day \cite{yahoo}. Trading during the day may have affected these numbers. 
The reason for the large number of such options is the round price at expiration and that this Friday in January was a unique one during the year whose associated options were available to buy and sell for two years. Two explanations for the closing price have been put forward: Manipulation by those who stood to gain by the expiration of those options \cite{manip0,manip1,manip2}, and legitimate stock trading activities that are used to offset (hedge) changes in option value \cite{hedge1,hedge2,hedge3,disc1,disc2}. 
Both explanations are based upon the recognition that the expiration of options has a potential influence on the price of the stock, especially close to their expiration dates. On such a date the precise closing price determines which options expire without value. Here we quantitatively investigate the explanation that the closing price is due to legitimate trading based upon hedging, and show that this explanation cannot account for the movement of the stock price.

Options are highly leveraged ``side-bets'' on stock price movements. Sellers and buyers gain or lose in a zero sum game betting on the price movements of stocks. The high leverage arises because the price of options
is more closely related to the change of stock price than the total stock price, so small bets can provide large gains if the seller or buyer anticipates the movement of the underlying stock price. 
The percent gain of an option is often between ten and one hundred times the percent gain of the underlying stock price. 
A ``call'' option is an offer to buy a stock at a pre-specified price on a particular day, the expiration day, and a ``put'' option is an offer to sell a stock at a pre-specified price on the expiration day. If the price will go up it is good to buy a ``call'' option or sell a ``put'' option, and vice versa if the price goes down. The importance of options to the stock market has increased in recent years, particularly since the frequency of options expiration dates was increased from monthly to weekly on July 1, 2010. 

According to traditional analyses, the option price is determined by the stock price which is itself determined by the fundamental value of the company for which the stock is issued \cite{blackscholes,merton}. However, it has been found in theoretical and empirical studies that options trading influences the price of stocks. The extent of this influence is not well understood, however, it has been shown that options can change stock price volatility \cite{frey,sircar,wilmott,clustering2}, even affecting the entire market \cite{wang2012}. Furthermore, on options expiration dates, it has been shown empirically that stock prices tend to aggregate near the expiration strike prices of the options more often than they would at random \cite{Krishnan2001,Avellaneda2003,clustering1}. This effect is called clustering or pinning. Thus, for example, since options strike prices sold for Apple are multiples of \$5, the expiration price is likely to be closer to such a strike price than expected by random price movements. It is now well established that closing prices prior to expiration dates are affected by options. However, there may be multiple mechanisms and the specific reason that this occurs in any one case or in general is not clear.

Given the volatility of the price of Apple stock the likelihood that the Friday, January 18,  2013, Apple Inc. stock (AAPL) close occurred at exactly \$500.00 per share by chance on this specific date is less than one in 1,000. Options effects on expiration date closing prices provide potential explanations for the special closing price. There are two explanations for this occurrence and there is some empirical evidence that they have occurred in other circumstances \cite{clustering1}. The first cites manipulation by those who sell the underlying stock to cause the options that they have sold to expire, resulting in no obligation on their part to fulfill the contracts. The second attributes price movements to the legitimate buying and selling of stock for the purpose of reducing risk (hedging) in corresponding options positions by large traders, often specific organizations that are considered ``market makers,'' who participate in the market to profit from executing many transactions rather than from stock price movements \cite{Krishnan2001,Avellaneda2003}. These effects have been suggested to give rise to price clustering at expiration and other price movements that result in maximum losses to those who buy options, the maximum pain theory (maxpain) \cite{maxpain,optionpain}. 

The first explanation is anchored in the belief \cite{PED2011_1,PED2011_2,PED2011_3}, supported by theoretical \cite{jarrow} and empirical \cite{clustering1,liu2009} evidence, that large traders take the future into their own hands through market manipulation.  According to this explanation, reliable profits can be made by buying or selling options and subsequently manipulating the underlying stock price. Rather than stock prices being an outcome of buying low and selling high, or even trend following, prices are driven in the direction that causes the desired option price changes. A widely publicized example occurred 20 years ago in the case of a purchase by a single trader of \$500 million of options on Venezuelan bonds from a broker-dealer, whose expiration value was subsequently subject to a trading war in which each side of the original options trade strove to move prices in the direction that would favor themselves \cite{bondfight1,bondfight2}. More generally, market manipulation has been analyzed and shown to occur \cite{aggarwal,misra}. The idea that markets may deviate temporarily but will eventually come into equilibrium does not apply to options because they expire at a particular date. Manipulation would intensify toward an expiration date, with distinct groups competing with each other based upon purchasing power to drive the price in the direction they need to make a profit.

The second explanation is based upon an understanding of the process of risk reduction by hedging \cite{Krishnan2001,Avellaneda2003,salpha1,Street2011}. A market maker sells or buys options depending on the demand from others.
However, the market maker may not want to carry the risk associated with the positions they accumulate. 
One of the standard strategies that may be used by market makers to reduce their risk is by offsetting gains or losses in the options by buying or selling the underlying stock. 
Thus, if a market maker sells options to buy a stock, they themselves will be responsible for delivering the stock at expiration and therefore they must hold the stock in case it increases in value. Similarly, if they sell options to sell a stock, they would hold a negative ``short'' position in the stock so that the effect of delivery of a stock to them at expiration will be offset as well. 
The amount of stock they have to own changes as time progresses toward expiration. This process of hedging options against the underlying stock price is called ``delta hedging.'' The rate of change of the value of the options as a function of price variation in the stock is called ``delta.'' The aggregate value of delta for the position is the amount of stock that a hedger must own to prevent a price movement from changing the value of the position. 

The case of stock prices moving toward strike prices by hedging activity occurs when the market maker has a net (positive) position of call and put options at a specific strike price near the price at which the stock is trading during the final hours of trading. By contrast, if the market maker has a constant net negative option position then the price will be driven away from the strike price by hedging. While it seems that market makers are more likely to be net sellers of options, there is evidence that under certain circumstances they end up as net buyers \cite{clustering2}. Suppose market makers are following a delta-hedging strategy. If they are holding call options, and the price is above the strike price, then they will hedge by selling the underlying stock. If they are holding put options, and the price is below the strike price, then they will hedge by buying the underlying stock (regardless of whether they also hold call options, because those options are due to expire valueless). If this is a sufficient percentage of the market activity, selling above the strike price, and buying below the strike price, will act as forces on the price, driving it toward the strike price from above or below, ``pinning" the price to the strike price approaching expiration. Strike prices that have more options associated with them would according to this theory have a stronger attraction, providing a tentative explanation of the close at \$500.00, whose round number and extended period of options availability resulted in a particularly large amount of options sold. 

We will show, however, that this picture of legitimate hedging does not apply to the price behavior on Friday, January 18 based upon a direct analysis of the price movement on that day. The price movement toward the strike price from below occurred at a time and with a rate of change that would have required delta-hedgers to respond by selling rather than buying, driving the price away from rather than toward the strike price. The price approached the strike price during the final hour of trading from below. Price movements themselves result in a need for hedgers to change their positions. During this period, the price movement toward the strike price was so fast that the hedgers would have sold rather than bought stock. This means that the force of hedging, to the extent that it occurred, would have opposed the price movement and cannot be used to explain the movement of the price toward its closing value.

\begin{figure}[t]
\refstepcounter{figref}\label{datprice}
\includegraphics[width=.75\textwidth]{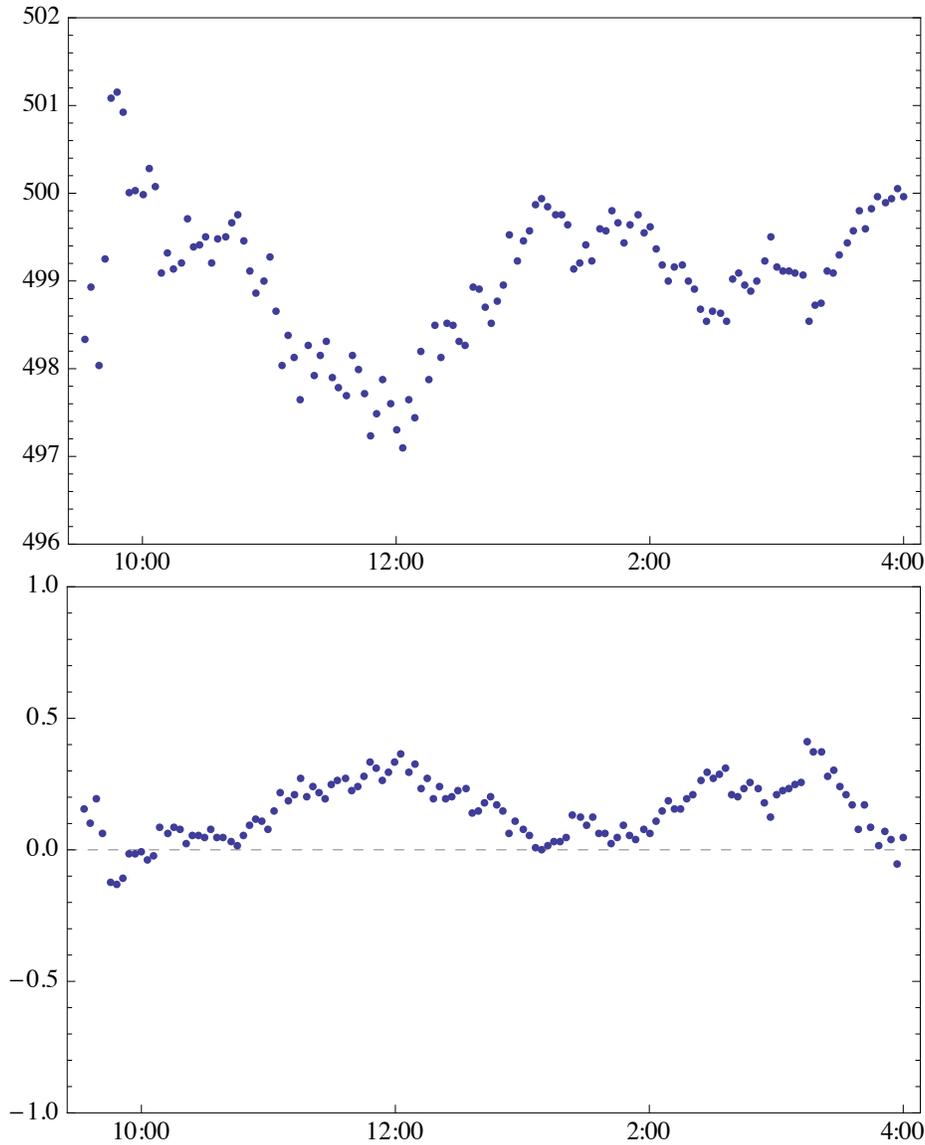}
\caption{ 
(A) Price of AAPL stock during January 18, 2013. (B) The amount of hedging needed by the owner of straddles (equal numbers of puts and calls) as determined by the stock prices. During the closing hour the amount of hedging needed decreases and the stock must be sold which would drive the price down, away from the strike price rather than toward it. The reason for the decrease in hedging is due to the time and rate of change of the prices toward the strike price. These results are counter to the claim that hedging is responsible for the closing strike price.}
\end{figure}

Figure \ref{datprice}A shows the actual AAPL stock price over the course of January 18, 2013, which oscillates before moving upwards toward the option strike price at \$500. Figure \ref{datprice}B shows the corresponding values of the amount of hedging needed to counter those prices based upon a calculation of delta, the rate of change of option values relative to the stock price \cite{blackscholes,merton}. For hedging to be responsible for the increasing price, the amount of stocks that are needed for hedging would have to increase, leading them to buy. Instead we see that the amount decreases over time as the price moves toward the strike price during the last hour of trading. The reason for the decrease is that the hedging fraction is smaller for prices nearer to the strike price at a particular time, while it increases over time at a given price. How much hedging is needed thus depends on both time and price. 

The size of the AAPL price movement at the time it took place is inconsistent with hedging. While the amount of hedging that is needed increases with time, overpowering this trend are movements of the amount of hedging toward zero, mirroring movements toward the strike price in the price data. This shows that the approach of the price to the strike price is faster than would be explained by pinning. Indeed, the price is approaching the strike price in spite of any hedging behavior that might be taking place by market makers, not because of it. Thus hedging activity cannot explain the approach of the stock price toward the strike price. 

We can construct a dynamic model of hedging based upon a quantitative description of delta hedging activity. We use a modified version of a model presented previously by Avellaneda and Lipkin \cite{Avellaneda2003}. The modification is necessary because they neglected to include one of two terms in the hedging activity. Specifically, they included the time dependence of the hedger's position but not its price dependence. 

The model is based upon two equations, one that characterizes the volume of trading that is performed by the hedgers and one that calculates the impact of that trading on market prices. Price changes in the market are proportional to the amount of buying and selling hedgers do according to price elasticity with large trades \cite{Avellaneda2003}:
\begin{equation}
\frac{\Delta S}{S}=EQ
\end{equation}
where $S$ is the current (spot) price, and $\Delta S/S$ is the relative change in spot price, which is  proportional to $Q$, the size of the excess supply ($Q<0$) or demand ($Q>0$) of the stock traded. The proportionality constant $E$ represents the price-demand elasticity of a given stock.

Consider the case in which the supply/demand $Q$ is due to market makers hedging their positions. In particular, we assume that the market makers own a net of $n$ straddles (both a call and a put) with strike price $K$ and expiration time $t_0$. Their hedging is proportional to the change in the delta of a straddle on the stock with respect to time multiplied by $n$. The delta, $\delta (S,\tau)$, is a function of the current price and the time until expiration,  $\tau=t_0-t$, given to first order by
\begin{equation}
\delta(S,\tau)=2N(d_1)-1
\end{equation}
where $N$ is the cumulative normal distribution function, and
\begin{equation}
d_1=\frac{1}{\sigma\sqrt{\tau}}\left(\ln\!\left(\frac{S}{K}\right)+\left(\mu+\frac{\sigma^2}{2}\right)\tau\right)
\end{equation}
where $\sigma$ is the implied volatility and $K$ is the strike price of the options.
Then we have:
\begin{equation}
\frac{\Delta S}{S}=-En\frac{d \delta(S,\tau)}{d t}\Delta t
\end{equation}
The derivative has two terms, one is the contribution directly due to time variation, i.e. the partial derivative with respect to time, 
\begin{equation}
\frac{\partial\delta(S,\tau)}{\partial t}=\frac{1}{\sqrt{2\pi}}e^{-\frac{d_1^2}{2}}\left(\frac{1}{\sigma\tau^{3/2}}\ln\!\left(\frac{S}{K}\right)-\frac{\mu+\frac{1}{2}\sigma^2}{
\sigma\sqrt{\tau}}\right)
\end{equation}
and the second is the derivative with respect to price times the price change with respect to time,
\begin{equation}
\frac{\partial\delta(S,\tau)}{\partial S}\frac{d S}{d t}=\frac{1}{\sqrt{2\pi}}e^{-\frac{d_1^2}{2}}\frac{2}{\sigma\sqrt{\tau}S}\frac{dS}{dt}.
\end{equation}
The original derivation by Avellaneda and Lipkin included only the first of these terms (we also corrected an errant factor of two in their formula). 
The total derivative is
\begin{equation}
\frac{d\delta(S,\tau)}{d t}=\frac{1}{\sqrt{2\pi}}e^{-\frac{d_1^2}{2}}\left(\frac{2}{\sigma\sqrt{\tau}S}\frac{d S}{d t}+\frac{1}{\sigma\tau^{3/2}}\ln\!\left(\frac{S}{K}\right)-\frac{\mu+\frac{1}{2}\sigma^2}{\sigma\sqrt{\tau}}\right) .
\end{equation}
Collecting terms, we obtain the differential equation ($\Delta t \to 0$):
\begin{equation}
\frac{dS}{dt}
=\frac{S\left[\mu+\frac{1}{2}\sigma^2-\frac{1}{\tau}\ln\!\left(\frac{S}{K}\right)\right]}{\frac{\sigma\sqrt{2\pi\tau}}{En}e^{\frac{1}{2}d_1^2}+2}
 \label{diffeq}
\end{equation}
We can make scaling substitutions of price, time, volatility and elasticity
\begin{equation}
\label{substitutions}
z=\frac{\ln\!\left(\frac{S}{K}\right)}{\sigma\sqrt{t_0}},\quad s=\frac{t}{t_0},\quad \alpha=\frac{\left(\mu+\frac{1}{2}\sigma^2\right)\sqrt{t_0}}{\sigma},\quad \beta=\frac{nE}{\sqrt{2\pi\sigma^2t_0}}
\end{equation}
to obtain $d_1=\frac{z}{\sqrt{1-s}}+\alpha\sqrt{1-s}$ and
\begin{equation}
\frac{dz}{ds}=\frac{\alpha-\frac{z}{1-s}}{\frac{\sqrt{1-s}}{\beta}e^{\frac{z^2}{2(1-s)}+\frac{1}{2}\alpha^2(1-s)+z\alpha}+2}
\end{equation}

\begin{figure}[t]
\refstepcounter{figref}\label{simprice}
\includegraphics[width=.75\textwidth]{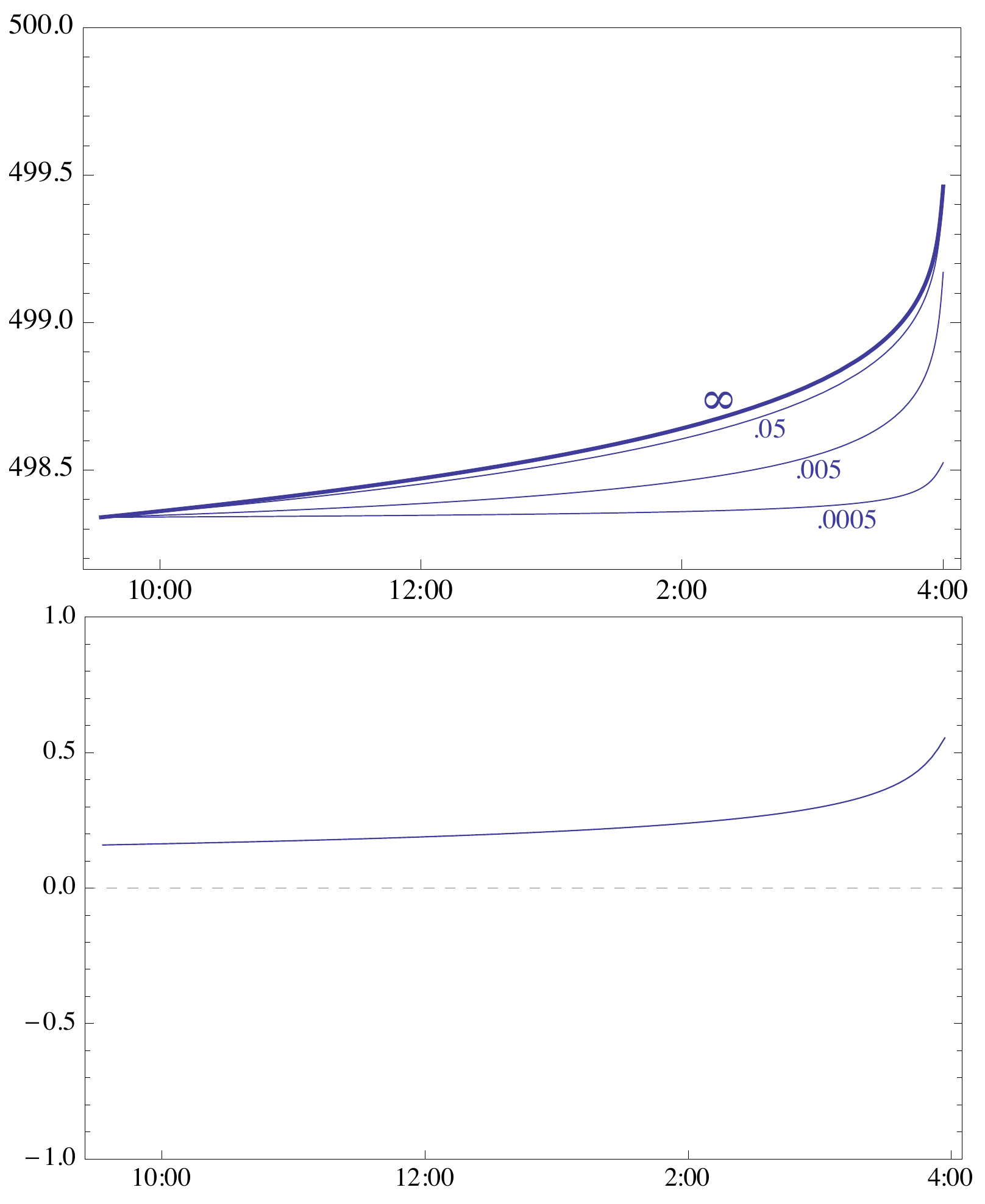}
\caption{(A) Simulation of the effect of hedging on prices until 3 minutes before 4:00 PM close at expiration. Curves are for different hedging impacts, $nE/\sqrt{2\pi}$, which saturates at a maximum, $\infty$, due to the price changes impact on hedging. (B) Time series for hedging based on the simulated price. Parameters are based on AAPL stocks on January 18, 2013. The price at the 10:00 AM open, \$498.34, strike price $K=\$500.00$, and the implied volatility $\sigma=1.102\times 10^{-3} /\sqrt{\min}$ \cite{bloomberg}.}
\end{figure}

Solving numerically, using parameters appropriate to the case of AAPL on January 18, 2013, we obtain the dynamics shown in Figure \ref{simprice}A. In Figure \ref{simprice}B, we show the the amount of hedging $N(d_1)$, calculated from the simulated price. As the expiration time approaches, hedging increases, driving the stock toward the strike price (as is expected from pinning). However, the approach toward the strike price is limited by a feedback from the price dependence of the hedging. The hedger that buys stock causing a price increase dynamically adjusts the amount of stock purchased as that same price increase leads to a smaller need for hedging. The maximal impact of hedgers is obtained by taking the limit as $E$ goes to infinity ($\beta \to \infty$)
\begin{equation}
\frac{dz}{ds}=\frac{1}{2}\left(\alpha+\frac{z}{s-1}\right)
\end{equation}
which can be solved analytically to yield:
\begin{equation}
z=-\alpha(s-1)+k \sqrt{s-1}
\end{equation}
where $k = \sqrt{1 - s_0} (z_0 - \alpha (s_0 - 1))$ is a constant. Note that in all cases the price behavior is very different from that found for AAPL on January 18 as price movements occur much closer to the closing. 

\begin{figure}[t]
\refstepcounter{figref}\label{simnoise}
\includegraphics[width=.75\textwidth]{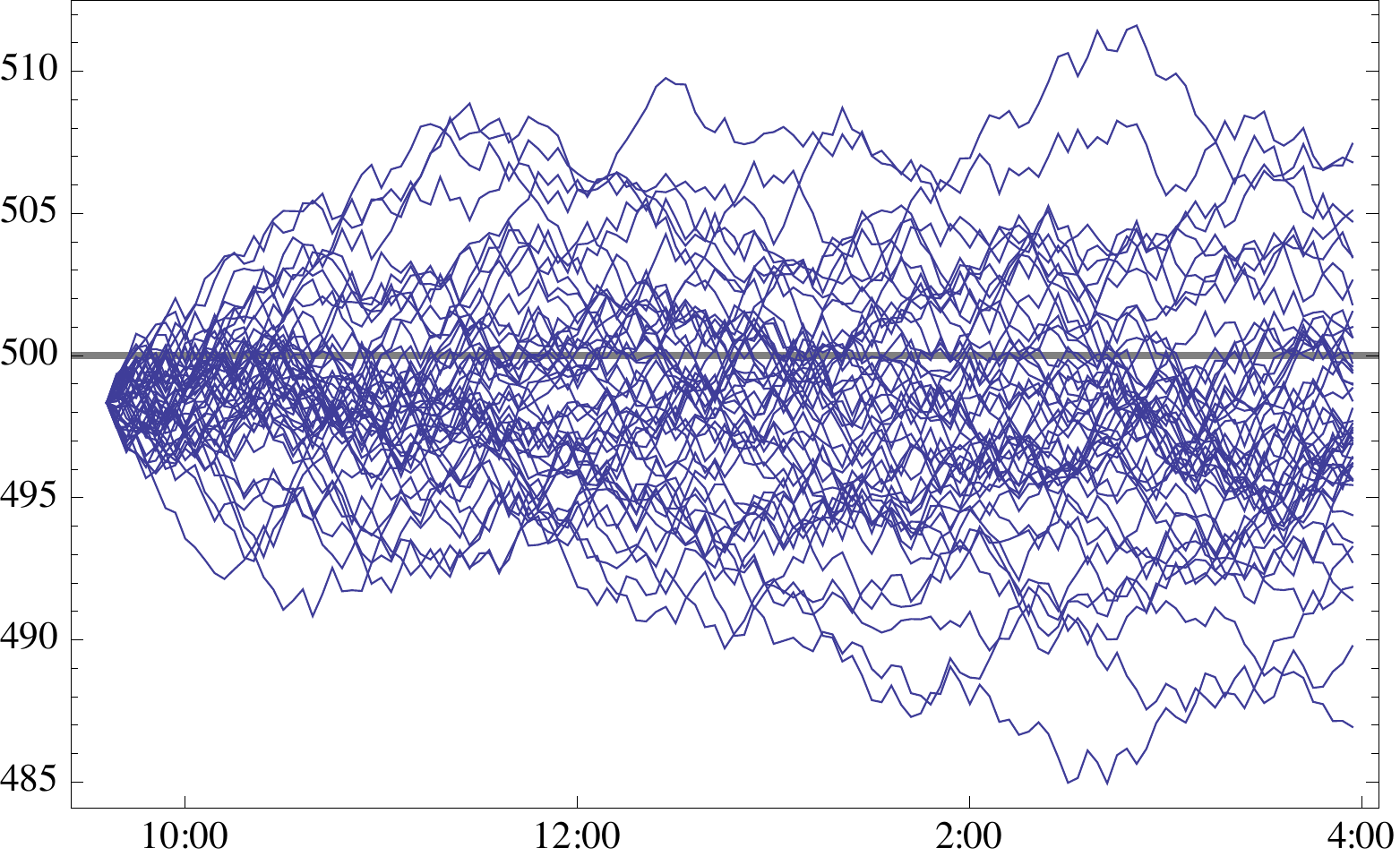}
\caption{Simulation of the effect of hedging on prices including noise. Fifty instances are shown. Parameters are based on AAPL stocks on January 18, 2013, and noise consistent with the average intraday volatility in January. The price at the 10:00 AM open, \$498.34, strike price $K=\$500.00$.}
\end{figure}

Random price movements due to other trading activity disrupt the process of hedger's influencing prices. Simulations with noise are shown in Figure \ref{simnoise}, with parameters appropriate to Apple on January 18, 2013, and using the average intraday volatility for January. The wide range of closing prices gives an estimate, with hedging, of the probability of the final price reaching its the strike price of less than $1$ per thousand based upon $5,\!000$ runs, even with the starting price as close as it was to $\$500.00$ on that day. The effect of hedging is weak compared to the underlying price volatility. 

The scenario we have analyzed is based upon the premise of a fixed number of options held by the hedger throughout the expiration date. There are two other scenarios that allow for a substantial number of options traded during the day of expiration. Option trading during the day may be done for two reasons. First is the usual opportunity for traders to buy and sell options, which may, but often does not, result in a major change in the number of open positions. The second is the possibility of the large number of pre-existing option positions being closed before expiration in order not to incur the cost of the subsequent ownership or sale of underlying stock.

The first  trading scenario does not provide a sufficient explanation for the close at $\$500.00$ as it would not be unique to the specific closing day or closing price. Any expiration day, i.e. any Friday, and any expiration price, i.e. multiple of $\$5.00$, would give rise to the same conditions. Such an explanation is therefore insufficient. Only one prior close at a strike price happened in 133 weeks since the introduction of the weekly options in July of 2010 \cite{yahoo}, and therefore would not explain the coincidence of special date and price on January 18. 

The second trading scenario would be based upon a different assumption about the hedging. Instead of the hedger having bought the options, the hedger would be assumed to have sold the options. Under these conditions traders may close their positions by selling the options on the expiration day, and the hedger would purchase the options and subsequently reduce its hedging position accordingly. The reduction in hedging would drive the price upwards if the price is below the strike price, nominally giving a reason for reaching the strike price. However, under these conditions, the underlying hedging would drive the price away from the strike price causing the strike price to be unstable, and the driving force away increases as the price moves toward the strike price, and as the time approaches the closing. On the other hand the impact of the sale of options is reduced as the price approaches the strike price because the level of hedging is smaller.  Moreover, a residual position held by the hedger at the end of the day would still result in the price being driven away from the strike price at closing. Simulations confirm there is no inherent reason for the closing price to be the strike price. This scenario does not, therefore, provide a robust explanation for closing at the strike price. 

It is worth noting that for the case of a hedger having sold options ($n$ negative), Eq.  \ref{diffeq} can be singular due to cancelation of the two terms in the denominator. This singularity corresponds to having insufficient stock trading volume to provide hedging for the outstanding options. Under such conditions the hedging activity will drive the price far away from the strike price resulting in an undefined, i.e. unstable, market price. 

In contrast to the hedging explanations, the motivation and mechanism of manipulation to a close at $\$500.00$ arises when large traders, who might be market makers, choose not to hedge, but rather to manipulate. For this case, a market maker would gain the most after selling more options than buying by causing them to expire without value. The mechanism is straightforward and is consistent with the price behavior on January 18: A sufficiently strong market maker or large trader can force convergence to the strike price by trading directly based upon the price rather than the option values, progressively confining it to a smaller range, until the close. Given the large amount of options outstanding, this transparency of the ``simple explanation'' led two commentators to predict the close \cite{manip0,manip1}, and to explain the motivations involved.

Our analysis suggests that it is possible to distinguish the role of hedging and manipulation on price movements at expiration. It also provides new evidence that ``side-bets'' on price movements create a mechanism for market manipulation and the associated financial gains. 

The existence of widely ignored evidence of market manipulation raises deeper questions about the oversight of markets. The Securities and Exchange Commission (SEC) has been criticized for lax oversight \cite{sec1,sec2,sec3,sec4}. 
Some attribute the behavior of the SEC to regulatory capture, a well known and long standing reality of regulatory agencies being influenced by the companies they are supposed to regulate \cite{wilson}; even a former top official at the SEC has commented on the state of capture of the commission \cite{sec3}. Given the size of economic benefits from such an arrangement, and the covert nature of the influence, it is well worthwhile to market participants to offer benefits to regulators such as monetary bribes or future job prospects. We separately provide a game theory analysis of regulatory capture showing that it is the economically rational outcome and therefore to be assumed unless the impact of regulation (or lack thereof) becomes transparent \cite{albino}.

In this context we can challenge the role of regulators and advocate for increased market transparency to allow improved public visibility. Large amounts of data can be made available for scrutiny for evidence of manipulation. Today, the stock holdings of large stock owners are made public. The current three-month intervals at which such holdings are revealed are insufficient. The original purpose of this public disclosure was to protect companies and their investors from insider trading \cite{SEC1934}. Today, however, an important means of manipulation is by performing large volumes of trading at a particular time, rather than through high levels of ownership and insider information. Since manipulation of stocks 
typically involves large transactions over a particular interval of time, especially a short one, such transactions should be made public. Moreover, position reporting should be extended to those who have sold short large amounts of the same stock after borrowing them, which is currently not included in public information. Transaction data should be independently made public to enable verification of the large transaction public reporting, which like the absence of illegal manipulation cannot otherwise be gauged. 

Detecting market manipulation of expiration date closing prices is not difficult given the right data, and that data is surely recorded. The pattern of trading of specific individuals or groups that buy and sell both options and the underlying stock can be found. Among the likely candidates for manipulation are the main market makers that provide liquidity to the options market. While these market makers are often large banks \cite{mmNYSE,mmNASDAQ}, the expectation that such financial institutions are beyond culpability has been undermined \cite{financialcrisis,mortgages1,mortgages2,overview,LIBOR,laundering,sanctions,auctions,HAMP,metals,cookingbooks,pensions}.  There is a direct financial motivation for such manipulation. While legitimate hedging reduces their risk from price movements, manipulation leads to immediate gains. 

Public disclosure of market makers and their activity may shed light on conflicts of interest and potential roles in market manipulation. It has been stated that the primary options market maker for AAPL is Goldman Sachs  \cite{maximumpain}, though this information is not widely discussed. We note that earlier in 2012, Goldman Sachs options investors publicly recommended an options trade that would have provided profits if AAPL did not rise significantly. This trade ended badly for those who adopted their recommendation, but profited the market maker that sold them those options  \cite{GSrecommends}. A second recommendation by Goldman Sachs in May 2013 to buy call options similarly resulted in losses for the buyers and gains for the market maker \cite{GSrecommends2}. These were the only two options purchase recommendations reported in the press during this period. 

What are the mechanisms for defeating market manipulation making use of options? First, corporations and investors should be advocates of proper market oversight, and the elimination of incentives and mechanisms of manipulation. Second, options trading should be reduced voluntarily, if not by regulator action, to reduce the benefits from this kind of manipulation. Third, manipulative price movements should be clearly identified, and explanations as legitimate hedging be denied, so that non-enforcement by the SEC is a matter of public dialog. Fourth, the possibility of class action suits against market makers that can be identified should be explored. Fifth, data should be made publicly available to increase the transparency of markets and expose manipulation as it occurs. A final, and encompassing, option is to pursue collective action of investors to stabilize markets and reduce the possibility of market manipulation. 

We thank Mark Kon for helpful discussions.

\end{document}